\documentclass[proof]{pasj00}   
\SetRunningHead{Prompt Emission of GRB100418A}{Imatani et al.}

\usepackage{times}
\usepackage{lscape}
\usepackage[dvips]{graphicx}
\usepackage{ulem}
\newcommand{\GRB}{GRB\,100418A}  
\newcommand{\XRTtime}{71\,sec}  
\newcommand{\EpSSC}{8.3\,keV}  
\newcommand{\red}{\textcolor{black}}
\newcommand{\redA}{\textcolor{black}}
\newcommand{\Dtime}{$31.6\pm1.6$} 

\begin{document}

\title{Soft X-ray Observation of the Prompt Emission of \GRB}
\author{
        Ritsuko \textsc{Imatani},\altaffilmark{1}
        Hiroshi \textsc{Tomida},\altaffilmark{2}
        Satoshi \textsc{Nakahira},\altaffilmark{3}
        Masashi \textsc{Kimura},\altaffilmark{3}
        Takanori \textsc{Sakamoto},\altaffilmark{4}
        Makoto \textsc{Arimoto},\altaffilmark{5}
        Yoshitaka \textsc{Morooka},\altaffilmark{6}
        Daisuke \textsc{Yonetoku},\altaffilmark{7}
        Nobuyuki \textsc{Kawai},\altaffilmark{5}
        and Hiroshi \textsc{Tsunemi},\altaffilmark{1}
}
\altaffiltext{1}{%
   Department of Earth and Space Science, Osaka University, 1-1 Machikaneyama-cho,  Toyonaka, Osaka 560-0043}
\altaffiltext{2}{%
   Institute of Space and Astronautical Science, 3-1-1, Yoshinodai, Sagamihara, Kanagawa 229-8510}
\altaffiltext{3}{%
   JEM Mission Operations and Integration Center, Human Spaceflight Technology Directorate,
    Japan Aerospace Exploration Agency, 2-1-1 Sengen, Tsukuba, Ibaraki 305-8505, Japan}
\altaffiltext{4}{%
   Department of Physics and Mathematics, Aoyama Gakuin University, 5-10-1 Fuchinobe, Chuo-ku, Sagamihara 252-5258}
\altaffiltext{5}{%
   Department of Physics, Tokyo Institute of Technology, 2-12-1 Ookayama, Meguro-ku, Tokyo 152-8551}
\altaffiltext{6}{%
Department of Applied Physics, Faculty of Engineering, University of Miyazaki, 1-1 Gakuen Kibanadai-Nishi, Miyazaki, Miyazaki 889-2192}
\altaffiltext{7}{%
College of Science and Engineering, School of Mathematics and Physics, Kanazawa University, Kakuma, Kanazawa, Ishikawa 920-1192}

\email{ritsuko@ess.sci.osaka-u.ac.jp}


\maketitle

\begin{abstract}
We have observed the prompt emission of \GRB\ from its beginning by the MAXI/SSC (0.7-7\,keV) on board the International Space Station followed by the {\it Swift}/XRT (0.3-10\,keV) observation.  The light curve can be fitted by a combination of a power law component and an exponential component (decay constant is \Dtime\ sec).  The X-ray spectrum is well expressed by the Band function with $E_{\rm p}\leq$\EpSSC.  This is the brightest GRB showing a very low value of $E_{\rm p}$.  It satisfies the Yonetoku-relation ($E_{\rm p}$-$L_{\rm p}$).  \redA{It is also consistent with the Amati relation ($E_{\rm p}$-$E_{\rm iso}$) in 2.5$\sigma$ level}.
 
\end{abstract}

\section{Introduction}
The observation of gamma-ray bursts (GRBs) with a very soft spectrum, so called X-ray flashes (XRFs), provides a unique information for understanding the nature of GRBs.  The bright GRBs observed by {\it Ginga} have values of $E_{\rm p}$ around a few keV \citep{1998ApJ...500..873S}, where $E_{\rm p}$ is a photon peak energy in the spectral form of $\nu$F$_{\nu}$ at the rest frame.  XRFs are observed both by {\it BeppoSAX} \citep{2004NuPhS.132..263H} and by {\it HETE-2} \citep{2005ApJ...629..311S}.  About one-third of GRBs observed by {\it HETE-2} are classified as XRFs.  Various theoretical models have been proposed for XRFs.  They include the external shock emission from low bulk Lorentz factor shells (e.g., \cite{1999ApJ...513..656D}), the off-axis jet viewing scenarios (e.g., \cite{2002ApJ...571L..31Y}), the X-ray emission from the hot cocoon of the GRB jets (e.g., \cite{2002ApJ...578..812M}), the very high redshift GRBs (\cite{2004NuPhS.132..263H}), the inhomogeneous jet model (e.g., \cite{2005ApJ...635..481T}) and the internal shock emission from the high bulk Lorentz factor shells (e.g., \cite{2005A&A...440..809B}).  After the launch of {\it Swift} (\cite{2004ApJ...611.1005G}), the observations of GRBs have been dramatically improved thanks to its rapid and accurate position information.  However, due to a relatively high energy threshold (around 15\,keV) of the Burst Alert Telescope (BAT; \cite{2005SSRv..120..143B}) on-board {\it Swift}, a very small number of XRFs are observed by {\it Swift} \citep{2008ApJ...679..570S}.  Furthermore, it is difficult to constrain $E_{\rm p}$ of XRFs based on the BAT data alone.  The observation of the prompt emission in X-ray is definitely needed to provide a crucial spectral information of XRFs.

{\it Swift} X-Ray Telescope (XRT; \cite{2005SSRv..120..165B}) has detected the X-ray emission of the early phase of the {\it Swift} GRBs (e.g., \cite{2006ApJ...642..389N, 2006ApJ...642..354Z}).  Some of the GRBs show steep decay components in the X-ray light curve (LC).  The origin of this steep decay component is believed to be a result of the delayed prompt emission from different viewing latitudes of the jet, so called a high latitude emission (e.g., \cite{2000ApJ...541L..51K}).  \citet{2006ApJ...647.1213O} studied the steep decay emission combining the BAT and the XRT LCs for 40 GRBs.  They found that the LC could be well described by an exponential decay relaxing into a power-law.  \citet{2010MNRAS.403.1296W} modeled a pulse profile, a spectral evolution and a high latitude emission of the prompt emission.  They found that the most of the steep decay component seen in the XRT could be described by the internal shock model.  One of the difficulties in those studies is to generate the composite LC in the fixed energy band.  They extrapolated the BAT data to the XRT energy band of 0.3-10\,keV.  Without knowing the spectral information in the X-ray band, the extrapolation may introduce a systematic error (see \cite{2007ApJ...669.1115S}).  The prompt emission observation in X-ray will be an ideal tool to connect the early GRB X-ray emission seen by the XRT.

When the BAT detects a precursor of the GRB, the XRT can observe its main part from the beginning.  \citet{2006A&A...456..917R} performed the XRT observation of the main part of GRB\,060124.  They reported that $E_{\rm p}$ was 636\,keV ($z$=2.297) with significant spectral evolution.  This is a very rare case that the XRT observation is carried out from its beginning of the main part. 


\red{
\GRB\ detected by the BAT belongs to a long GRB that is believed to be the death of massive stars based on the associations with supernovae (SN).  The XRT started the observation \XRTtime\, after the BAT trigger time \citep{2011ApJ...727..132M}.  The optical afterglow was also detected by the Ultra Violet Optical Telescope (UVOT;\cite{2005SSRv..120...95R}) on-board {\it Swift} while it reached the maximum brightness several hours after the burst in spite that the typical GRBs reach the maximum tens of seconds after the burst \citep{2008MNRAS.387..497P}.  A precise redshift z=0.6239 was determined by the follow-up observation \citep{2011AN....332..297D} showing that it was relatively a nearby source as the GRB.  This was also detected by the pre-ALMA observation being the third brightest burst in the mm/submm range \citep{2012A&A...538A..44D}.  \citet{2012PASJ...64..115N} could not identify any SN feature using FOCAS on the Subaru telescope.   The upper limit of the absolute magnitude was comparable to the faintest type Ic SN.
}
  Since \GRB\ occurred in the field of view (FOV) of the MAXI SSC and stayed for about 50\,sec, we observed its prompt emission prior to the XRT observation in the X-ray band (0.7-7\,keV).   In this paper, we report the prompt emission of \GRB.

\section{Observation}

MAXI \citep{MAXImission2009PASJ...61..999M} onboard the International Space Station is an all-sky X-ray monitor that consists of an array of gas proportional counters (GSC, \cite{2011PASJ...63S.623M}) covering the energy range of 2-40\,keV and an array of X-ray CCD (SSC, \cite{tsunemi2010PASJ...62.1371T, 2011PASJ...63..397T}) covering the energy range of 0.5-12\,keV.  \GRB\ occurred when the MAXI was close to the high background region (latitude was $-51^\circ$ in the South Pacific Ocean) where GSC was already turned off for safety.
 

The SSC consists of two identical cameras, SSC-H and SSC-Z, each has $2\times 8$ CCD array with different FOV, \red{a horizontal view (SSC-H) and a zenith view (SSC-Z).  The SSC scans the sky along a large circle of the ISS orbit in every 90\,min.}  It has a fan beam FOV of $1.^{\circ}5 \times 90^\circ$ (full width at half maximum, FWHM) where CCD functions as one-dimensional imager so that we can localize the source.  We usually read out 16 CCDs at every 6\,sec.  Two CCDs detected \GRB\ for an integration time of 6\,sec with 3\,sec time difference.  The maximum on-source area of the SSC is 1.35\,cm$^2$.

\red{MAXI/SSC monitors X-ray sources every day just as MAXI/GSC does.  We are monitoring the Crab nebula for its point spread function (PSF) of which uncertainty is about 5\%.  The charged particles hitting on the copper-made collimator generate Cu-K lines (8.04\,keV and 8.94\,keV) on all the CCDs.  They reduce the source detection efficiency at high energy that limits our effective energy range up to 7\,keV.  Calibration sources of $^{55}$Fe irradiate X-rays onto two CCDs out of 16.  The Cygnus Loop is also a strong source below 2\,keV \citep{2013PASJ...65...14K} that helps us to monitor the efficiency of CCD at low energy.  Due to the thermal noise development, we find that the effective energy range is above 0.7\,keV at the time of observation of \GRB.  In this way, we estimate the systematic uncertainty of the SSC spectral parameters for Crab-like source to be less than 10\% \redA{\citep{Kimura2012PhD} }.  These systematic uncertainties are less than the statistical uncertainties for \GRB.}

The BAT trigger time of \GRB\ was 2010/04/18,\,\,21:10:08 (UT) that was 2\,sec after the source got into the FOV of the \red{SSC-Z}.  The source was in the FOV for about 50\,sec.  Due to the program timer of the MAXI, the SSC got into the idle mode (bias voltage of the CCD was set to zero) at 21:10:53, just before the source left the FOV.  


\begin{figure} [htbp]
   \centering
   \includegraphics[width = 10cm ]{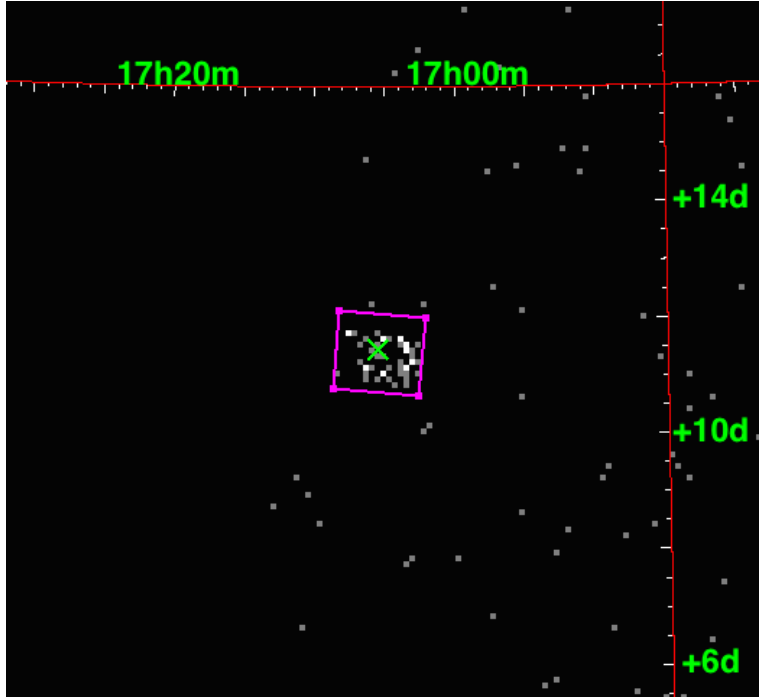}
   \caption{\GRB\ image obtained by the SSC-Z (0.7-7\,keV).  The green X mark indicates the GRB location ($17^h 05^m 25.^s8, +11^\circ 27^\prime 26.^{\prime \prime}8$, J2000) measured by the XRT.  The burst occurred when the source got into the FOV.  \red{The magenta rectangle corresponds to the SSC PSF (FWHM) for the \GRB.}   
}
   \label{sscimg}
\end{figure}

\section{Analysis and Results}
\subsection{Light curve (LC)}



\begin{figure} 
\begin{center}
\includegraphics[width = 10cm,angle=0]{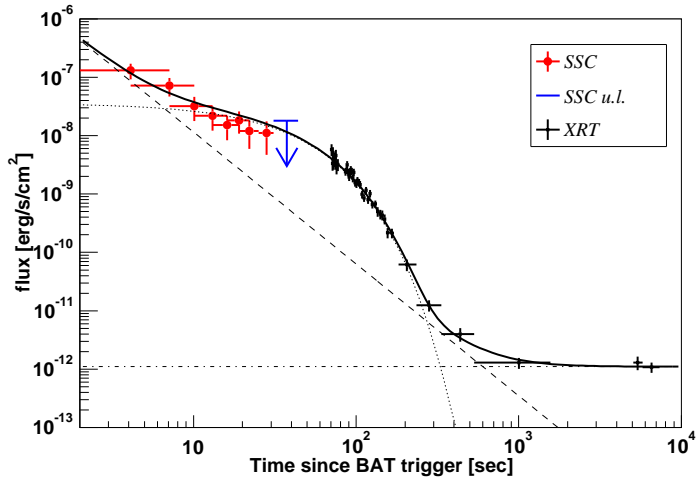} 
    \caption{\GRB\ Light Curve (0.7-7\,keV) : SSC data points (red) are Time$\leq$ 50\,sec and XRT data points (black) are \XRTtime\, $\leq$ Time.  Data from two CCD chips are plotted (see in text).  Mark in blue represents the upper limit of the SSC data.  \red{The solid line shows the best fit model LC while} three components are individually plotted; a power law component (dashed line), an exponential component (dotted line) and a constant component (dotted/dashed line).
 }
   \label{jointLC}
 \end{center}  
\end{figure}

As the apparent peak intensity of the \GRB\ was an order of magnitude higher than that of the Crab nebula, we collected 50 photons with background free.  Figure \ref{sscimg} shows the X-ray image where 50 photons form PSF of the SSC-Z ($1.5^\circ\times1.4^\circ$).  \red{The cross in the figure shows the position of \GRB, that is shifted about 0.3$^\circ$ from the center of gravity of photons.  This is due to the fact that the source intensity rapidly varies during the SSC-Z observation.} We select the energy range of 0.7-7\,keV and remove hot/flickering pixels.  \red{We collect 135\,photons in the $20^\circ\times20^\circ$ region centered on \GRB.  Among them, 50\,photons are in the SSC PSF.  With taking into account the fact that the SSC is turned off after the \GRB\ moved out of the PSF, the expected background count rate is 0.9\,photons/PSF.}
We correct the count rate by using the average collimator response for each 6\,sec.
We extract all the {\it Swift} data of \GRB\ from the web site (\verb|http://www.swift.ac.uk/burst_analyser/00419797/|).  Figure \ref{jointLC} shows the LC of \GRB\ obtained both by the SSC (Time $\leq$ 50\,sec) and by the XRT (\XRTtime $\leq$ Time).  In this LC, we calculate the flux (0.7-7\,keV) from the XRT data to fit the SSC energy range using the {\it WebPIMMS} ver.4.7d in {\it HEASARC}.  We see that the LC does not follow a simple power law of time.  
\red{Instead, we employ an exponential with a constant, which gives us C-statistics value (C-stat) of 77.7 \newline with 49 \redA{degrees} of freedom (dof).  We add a power law component, which gives us C-stat of 50.4 with 47 dof.  The final model is given in the equation below,}

\begin{eqnarray*}
{\rm flux} = K_{\rm e} \exp(-\frac{t}{w}) + K_{\rm p} t^{-a} + K_{\rm c} \nonumber
\end{eqnarray*}
where $t$ is the time after the BAT trigger, $w$ is a time constant, $a$ is a temporal index and $K_{\rm e}$, $K_{\rm p}$, $K_{\rm c}$ are normalizations .  The solid line in figure \ref{jointLC} is the best fit curve of which the parameters are given in table \ref{LCpara} \red{where errors are given in 90\% confidence based on the C-statistics.}

\begin{table}[hbt]
\caption{\sc Best fit values for the LC} 
\label{light_curve}
\begin{center}
\begin{tabular}{ccccc} \hline\hline
$K_{\rm e}$ & $K_{\rm p}$ & $K_{\rm c}$ & $w$ & $a$ \\
\hline
\multicolumn{3}{c}{ {\rm erg/cm}$^2${\rm /sec} } & {\rm sec} & \\ 
\hline\hline
$(3.8^{+0.7}_{-0.6}) \times 10^{-8}$ & $(1.6^{+1.6}_{-1.0} )\times 10^{-6}$ &$(1.1\pm 0.3) \times 10^{-12}$ &\Dtime &$2.26^{+0.21}_{-0.19}$\\
\hline
\end{tabular}\\
\small{Errors (90\% confidence) obtained with C-statistics}
\end{center}
\label{LCpara}
\end{table}

\begin{figure} 
\begin{center}
\includegraphics[width = 10cm ]{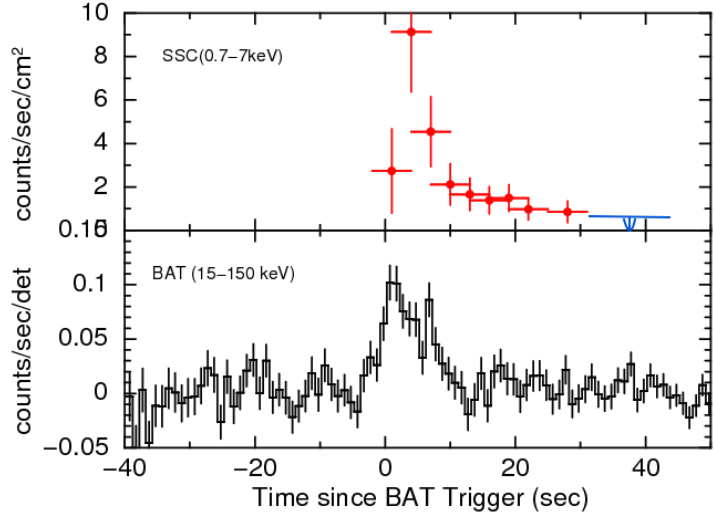}

    \caption{Prompt emission Light Curve of \GRB, (top) SSC (0.7-7\,keV), (bottom) BAT (15-150\,keV).  Data from two CCD chips are plotted (see in text).  The SSC detects no photon at 30\,sec$\leq$Time where we set the upper limit of 90\% confidence.   
}
   \label{promptLC}
 \end{center}  
\end{figure}

Figure \ref{promptLC} shows the \GRB\  LC around the BAT trigger time.  Data points of the SSC come from two CCD chips.  Each integrates data for 6\,sec with 3\,sec difference in integration start time.
The LC in the BAT energy band shows its maximum a little earlier  than that of the SSC.  This is consistent with the spectral lag usually seen in the prompt emission of GRBs \citep{1996ApJ...459..393N}.  The BAT LC shows two peaks whereas the SSC LC does not, probably due to the low time resolution. 

\subsection{Spectrum}

\begin{figure}
\centering
 \includegraphics[width=10cm]{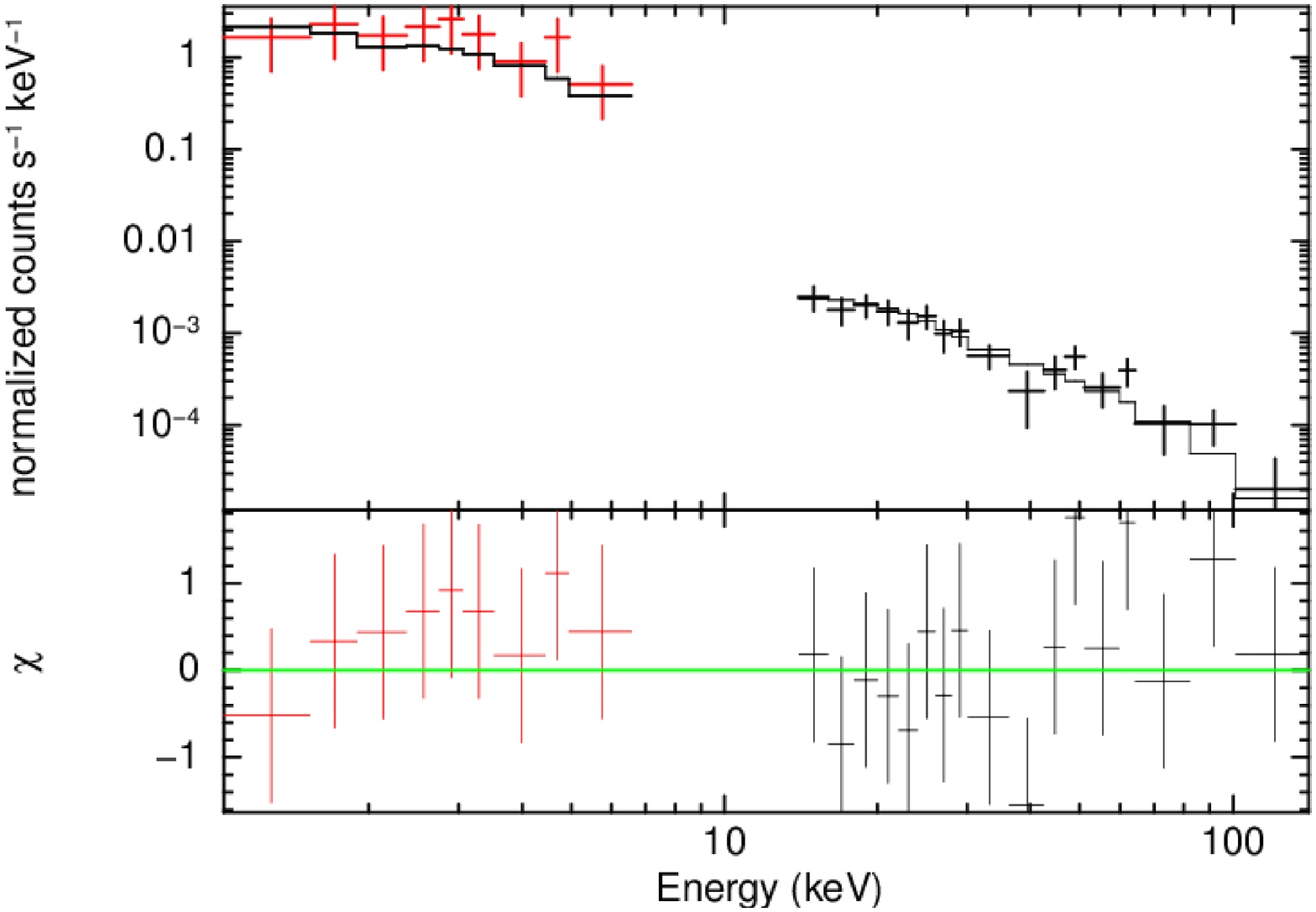}
\caption{Wide band X-ray spectrum of \GRB\ at the prompt emission, SSC ($E\leq$\,7\,keV) and BAT (15\,keV\,$\leq E$).  The best fit curve \redA{of the Band function} is shown in solid line in the top panel and residuals are in the bottom panel.  
}
\label{0-18Sp}
\end{figure}

In the prompt emission of \GRB, the BAT LC has a $T_{90}$ (the time interval containing 90\% of the flux) value of $8\pm 2$\,sec \citep{2011ApJ...727..132M}.  Therefore, most of the photons arrive in the first 10\,sec.  We employ the HEAsoft version 6.16 and CALDB version v4.3.1 in the BAT data analysis.  
\red{We collect SSC and BAT data for 18\,sec, starting 2\,sec before the BAT trigger time.  The time span is selected by the SSC read-out time.}  
Figure \ref{0-18Sp} shows the spectrum of \GRB, where SSC data are red crosses and BAT data are black crosses.  The spectrum of the BAT (15-150\,keV) can be fitted with a single power law of photon index $2.16\pm 0.25$ \citep{2011ApJ...727..132M} that clearly indicates that $E_{\rm p}$ lies below or near the low-energy threshold of the BAT.   \red{In our spectral fitting, 
\redA{we fix a cross-normalization factor between SSC and XRT to be unity.  We also fix the absorption} features to the values derived by the XRT : the Galactic absorption feature, $N_{\rm H}$(galactic), is fixed to the value of 4.8$\times 10^{20}$\,cm$^{-2}$ and the intrinsic $N_{\rm H}$(intrinsic) is fixed to the value of 3.3$\times 10^{21}$\,cm$^{-2}$\citep{2011ApJ...727..132M}.  The spectrum of the SSC can be fitted with a single power law with a photon index $1.3\pm 0.6$ where C-stat is 4.02 with 7 dof.  Then we combine these two spectra and fit them with a single power law model.  We obtain that the best fit value of the photon index is 2.1$^{+0.1}_{-0.2}$ where C-stat is 19.1 with 24 dof in table\,\ref{SPpara}.}  These results also indicate that the value of $E_{\rm p}$ must be below the BAT energy range.



We fit the data by using the Band function \citep{1993ApJ...413..281B}.  The low energy (below $E_{\rm p}$) photon index, $\alpha$, is typically $-1$ while that of the high energy (above $E_{\rm p}$), $\beta$, is about $-2.3$ \citep{2006ApJS..166..298K}.  Since $E_{\rm p}$ must be very low, we fix $\alpha$ to be $-1$ due to the insufficient data points while $\beta$ is left free.  The solid line in the figure is the best fit curve.  We also calculate the one-second peak luminosity, $L_{\rm p}$ and the isotropic equivalent energy, $E_{\rm iso}$ in the 1\,-\,$10^4$\,keV energy range with the assumption of the cosmological parameters of $\Omega_{\rm M} = 0.3$, $\Omega_{\rm \Lambda} = 0.7$, $H_0 = 70\,$km/sec/Mpc.  The results are summarized in table \ref{SPpara} \red{where the errors are 90\% confidence level based on the C-statistics.}

As we obtain the small value of the upper limit of $E_{\rm p}$, we check our fitting assumption.  If we set $\alpha$=$-0.5$ and $-1.5$, we find the upper limits of $E_{\rm p}$ to be \red{7.6\,keV and 9.2\,keV where the most plausible values are 4-4.5\,keV.} Therefore, we expect that the small value of the upper limit of $E_{\rm p}$ is very reliable.


\red{\citet{2011ApJ...727..132M} reported that they obtained a steep photon index ($4.33^{+0.28}_{-0.25}$) for the data between 84\,sec and 174\,sec after the BAT trigger time.  This is in stark contrast to our result for the very prompt emission.  We divide our data into two periods (the first 12\,sec and the rest) so that each period contains a similar number of photons detected.  Then we calculate the hardness ratio, R, (photon number ratio between the 0.7-4\,keV band and the 4-7\,keV band).  We obtain R=4.2$\pm$2.1 (1$\sigma$) and R=1.0$\pm$0.4 (1$\sigma$) for two periods, suggesting a spectral softening.  However, it is not clear from the statistical point of view.  Therefore, we can say that the rapid softening should have occurred before the XRT observation.}   
  
\begin{table}[hbt]
\caption{\sc Best fit values for the X-ray prompt emission spectral parameters (SSC+BAT and SSC)} 
\begin{center}
Power law function\\
\begin{tabular}{ccccccc} \hline\hline
$N_{\rm H}$(galactic)(fix) & $N_{\rm H}$(intrinsic)(fix) & z(fix)&\multicolumn{2}{c}{\underline{~~~~~~~~~~SSC~~~~~~~~~~}}&\multicolumn{2}{c}{\underline{~~~~SSC+BAT~~~~}}\\
	H/cm$^2$ &  H/cm$^2$& &  $\Gamma$ & C-stat/dof &  $\Gamma$ & C-stat/dof \\
	\hline
$4.8\times 10^{20}$ &$3.3\times 10^{21}$ & 0.6239 & 1.3$\pm$0.6 &  4.02 / 7 & $2.1^{+0.1}_{-0.2}$ &  19.1 / 24\\
\hline
\end{tabular}\vspace{6mm}\\
Band function\\
\begin{tabular}{cccccc} \hline\hline
 $\alpha$(fix) & $\beta$ & $E_{\rm p}$ & $E_{\rm iso}$ & $L_{\rm p}$ & C-stat/dof \\
  &  & keV & erg & erg/sec & \\
\hline
$-1$ & $-2.4^{+0.2}_{-0.3}$ & $\leq$ \EpSSC & $(1.7\pm 0.3)\times 10^{51}$ & $(1.4^{+0.6}_{-0.4})\times 10^{50}$ & 16.6/23\\
\hline
\end{tabular}\\
\small{$N_{\rm H}$(galactic), $N_{\rm H}$(intrinsic) and $z$ are the same to those in power law fit.\\$E_{\rm p}$ is the photon energy at the rest frame.}
\end{center}
\label{SPpara}
\end{table}

\section{Discussion and summary}

We have detected the prompt emission of \GRB\ in the soft X-ray energy range (0.7-7\,keV) before the XRT started the observation \XRTtime\, after the BAT trigger time.  \citet{2011ApJ...727..132M} reported that the prompt emission of the LC of the XRT could be expressed by a power law of time with an index of $-4.19$.   By adding the SSC data to the XRT data, we find that the prompt emission LC of \GRB\ in soft X-ray is well fitted not by a power law but by a combination of a power law and an exponential decay.  The exponential component dominates the emission between 7\,sec and 300\,sec after the BAT trigger time.  Then the flat component becomes dominant.


The LC of the prompt emission of GRBs have been reported as being  well described by an exponential decay (BAT LC) relaxing into a power law (XRT LC) (\cite{2006ApJ...647.1213O}).  \red{\citet{2007ApJ...669.1115S} investigated the rapid decay phase of the prompt emission for various GRBs and confirmed that the prompt emission LC could be expressed by a combination of a power law with an exponential decay.}  However, in these analyses, the LC of the prompt emission in the XRT energy range must be extrapolated from the BAT LC.  Our analysis of \GRB\ combining the SSC and the XRT data confirms those previous claims without extrapolation of data points beyond the observed energy range.

\begin{figure}
\centering
\includegraphics[width=7.5cm]{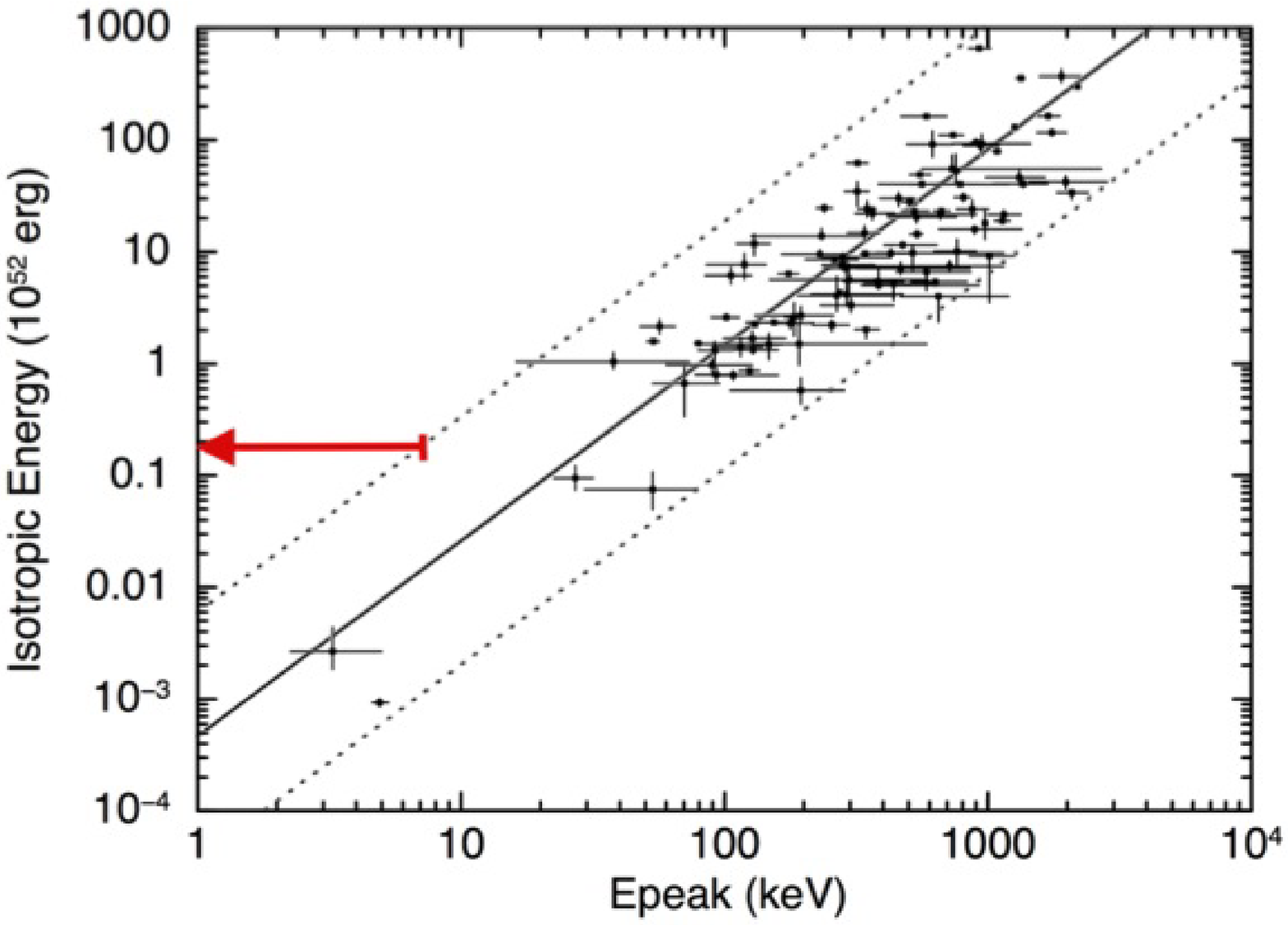}
\includegraphics[width=7.5cm]{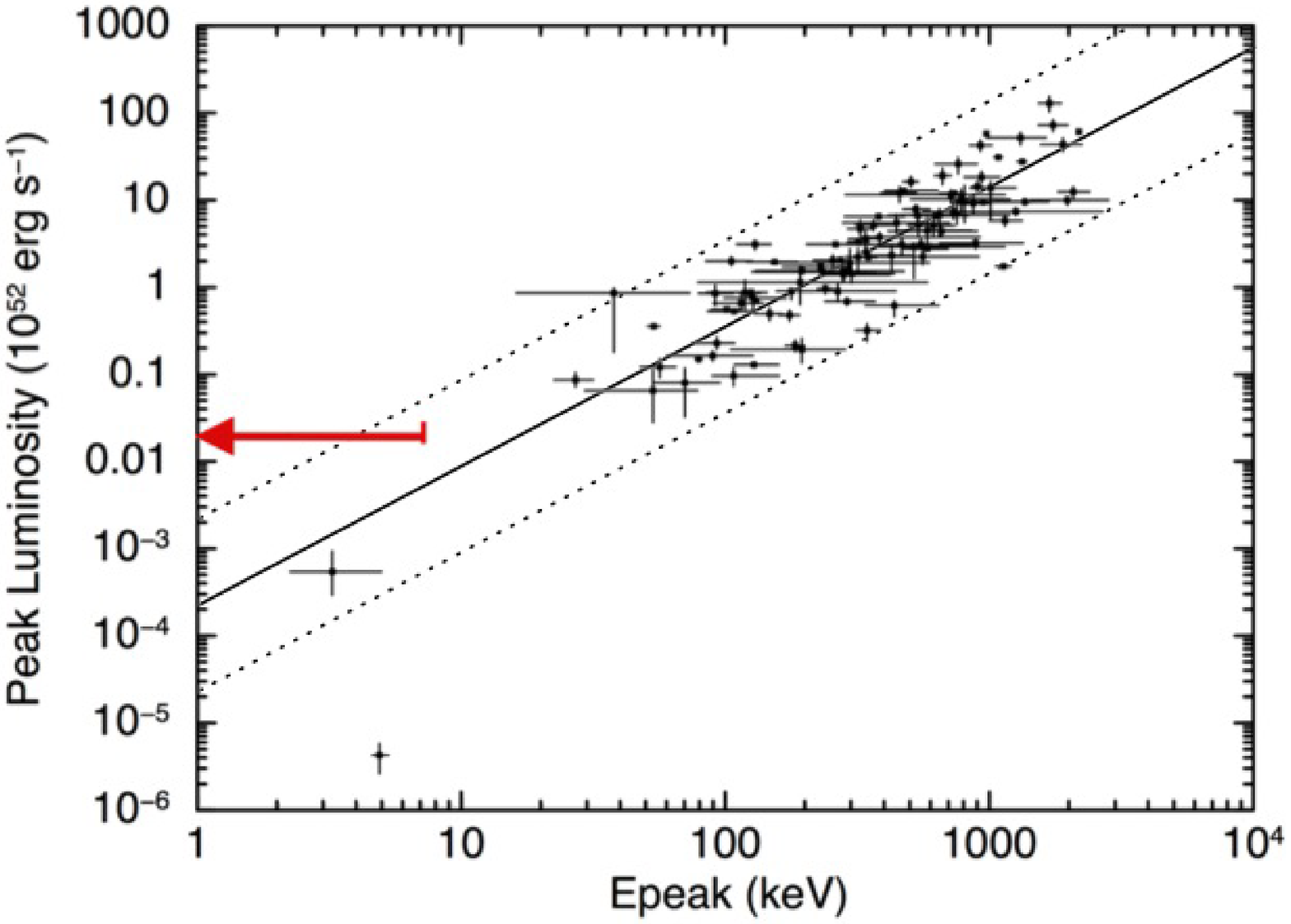}
\caption{Results of \GRB\ are plotted on the Amati-relation (left) and the Yonetoku-relation (right) \red{where dotted lines indicate the 3\,$\sigma$ systematic error regions \citep{2010PASJ...62.1495Y}}.  Red arrows indicate the results of \GRB.
}
\label{Amati_Yonetoku}
\end{figure}

The values of $E_{\rm p}$ of GRBs scatter in a wide energy range, from a few tens of keV to a few MeV \citep{2010PASJ...62.1495Y}.  Furthermore, there are two well known correlations: one is between $E_{\rm p}$ and $E_{\rm iso}$ (Amati relation: \cite{2002A&A...390...81A}) and the other is between $E_{\rm p}$ and $L_{\rm p}$ (Yonetoku relation: \cite{2004ApJ...609..935Y}).  These are studied well for long GRBs.  Recently, they are also applicable to the short GRBs (\cite{2012ApJ...750...88Z, 2013MNRAS.431.1398T}).  There are very few GRBs reported that have $E_{\rm p}\leq $10\,keV because it is very difficult to observe the prompt phase of GRB below 10\,keV by using the {\it Swift} data alone.


We obtain a very low value of the upper limit of $E_{\rm p}$ of \GRB\ in table \ref{SPpara}.  We plot the results of \GRB\ on the Amati-relation ($E_{\rm p}$-$E_{\rm iso}$) and the Yonetoku-relation ($E_{\rm p}$-$L_{\rm p}$) in figure \ref{Amati_Yonetoku} \red{where dotted lines indicate the 3\,$\sigma$ systematic error regions \citep{2010PASJ...62.1495Y}.  We notice that there are very few data points showing that $E_{\rm p}$ is below 10keV.  \redA{The upper limit point of the \GRB\ satisfies the Yonetoku-relation.  In the Amati-relation, it lies in the consistent level of 2.5\,$\sigma$ with taking into account the fact that we set the 90\% upper limit and that the Amati-relation in figure \ref{Amati_Yonetoku} shows 3$\sigma$ systematic error region.}  These relations may not be applicable at low energy $E_{\rm p}$.  It may be due to the fact that there are very few data points in $E_{\rm p}\leq $10\,keV region.  We need more data points whether or not these relations are really applicable at low energy region.} 

We find that the upper limit of $E_{\rm p}$ of \GRB\ is one of the lowest value actually obtained.  Other than \GRB ($E_{\rm p}\leq$ \EpSSC), there are only two GRBs showing its $E_{\rm p} \leq$ 10\,keV.  They are XRF\,020903 ($E_{\rm p}=3.3^{+1.8}_{-1.0}$\,keV) and GRB\,060218 ($E_{\rm p}=5.1\pm0.3$\,keV).  XRF\,020903 is an extremely soft X-ray flash \citep{2004ApJ...602..875S}.  In fact, there is no photon above 10\,keV.  Since the data points of XRF\,020903 satisfies both relations, they claim that the X-Ray Flash, the X-ray-rich GRB, belongs to the same class of classical GRBs.  \citet{2006Natur.442.1008C} analyzed GRB\,060218 classified as the Low Luminosity GRB (LL\,GRB) that was also unusual, showing a long $T_{90}$ of 2.1\,ks.  Its spectrum contained a thermal component as well as a power law component.  They also argued that this burst was dominated by a shock breakout.  The typical GRB emission is dominated by normal jet and afterglow components, therefore, LL\,GRB may show a different feature.  It may be the reason that GRB\,060218 satisfies the Amati-relation while it does not the Yonetoku-relation.  On the other hand, \GRB\ satisfies the Yonetoku-relation.  \redA{It is also consistent with the Amati-relation in 2.5$\sigma$ level}. Therefore, we need collect more samples to study the GRBs showing $E_{\rm p}\leq $10\,keV region.

\vspace{4mm}
In summary, we have observed the prompt emission of \GRB\ in the SSC energy range (0.7-7\,keV) from the very beginning of the BAT trigger time.  The LC is expressed not in a power law decay but in a combination of a power law and an exponential decay with a decay constant of \Dtime\,sec.  The X-ray spectrum shows $E_{\rm p} \leq$ \EpSSC.  \redA{It satisfies the Yonetoku-relation.  It is also consistent with the Amati-relation in 2.5$\sigma$ level}.  \GRB\ is a long GRB having a very low value of $E_{\rm p}$ reported.


\section*{Acknowledgment}

We thank all members of the MAXI operation and calibration teams. This work is supported by Japan Society for the Promotion of Science (JSPS) KAKENHI Grant Number 23000004, 24103002.  

\bibliographystyle{apj}
\bibliography{reference}


\end{document}